\documentclass[lettersize,journal]{IEEEtran}
\usepackage{amsmath,amsfonts}
\usepackage{algorithmicx}
\usepackage{algorithm}
\usepackage{algpseudocode}
\usepackage{changes}

\usepackage{array}
\usepackage[caption=false,font=normalsize,labelfont=sf,textfont=sf]{subfig}
\usepackage{textcomp}
\usepackage{stfloats}
\usepackage{url}
\usepackage{verbatim}
\usepackage{graphicx}
\usepackage{cite}
\usepackage{xcolor}
\usepackage{color}
\usepackage{array, makecell}
\usepackage{amssymb}
\begin{document}

\title{Soft Error Reliability Analysis of Vision Transformers}

\author{Xinghua Xue,
        Cheng Liu, ~\IEEEmembership{Member,~IEEE},
        Ying Wang, ~\IEEEmembership{Member,~IEEE},
        Bing Yang,
        Tao Luo, ~\IEEEmembership{Member,~IEEE},\\
        Lei Zhang,
        Huawei Li, ~\IEEEmembership{Senior Member,~IEEE},
        Xiaowei Li,~\IEEEmembership{Senior Member,~IEEE}

\thanks{Manuscript received April 25, 2023; revised August 17, 2023; accepted September 06, 2023. Date of publication xx xxx 2023; date of current version September 16, 2023. This work is in part supported by the National Natural Science Foundation of China under Grant 62174162. (Corresponding author: Cheng Liu.)}
\thanks{Xinghua Xue, Cheng Liu, Ying Wang, Lei Zhang, Huawei Li, and Xiaowei Li are with both State Key Lab of Processors, Institute of Computing Technology, Chinese Academy of Sciences, Beijing 100190, China, and University of Chinese Academy of Sciences, Beijing 100190, China. Bing Yang is with Department of Computer Science and Technology, Harbin University of Science of Technology, Harbin 150006, China. Tao Luo is with Institute of High Performance Computing, A*STAR, Singapore. (e-mail: \{xuexinghua, liucheng\}@ict.ac.cn)}
}

\markboth{IEEE TRANSACTIONS ON VERY LARGE SCALE INTEGRA TION (VLSI) SYSTEMS, VOL. XX, NO. XX, XXX 2023}%
{Xinghua Xue \MakeLowercase{\textit{et al.}}: Bare Demo of IEEEtran.cls for IEEE Journals}

\maketitle

\begin{abstract}
Vision Transformers (ViTs) that leverage self-attention mechanism have shown superior performance on many classical vision tasks compared to convolutional neural networks (CNNs) and gain increasing popularity recently. Existing ViTs works mainly optimize performance and accuracy, but ViTs reliability issues induced by soft errors in large-scale VLSI designs have generally been overlooked. In this work, we mainly study the reliability of ViTs and investigate the vulnerability from different architecture granularities ranging from models, layers, modules, and patches for the first time. The investigation reveals that ViTs with the self-attention mechanism are generally more resilient on linear computing including general matrix-matrix multiplication (GEMM) and full connection (FC) and show a relatively even vulnerability distribution across the patches. ViTs involve more fragile non-linear computing such as softmax and GELU compared to typical CNNs. With the above observations, we propose a lightweight block-wise algorithm-based fault tolerance (LB-ABFT) approach to protect the linear computing implemented with distinct sizes of GEMM and apply a range-based protection scheme to mitigate soft errors in non-linear computing. According to our experiments, the proposed fault-tolerant approaches enhance ViTs accuracy significantly with minor computing overhead in presence of various soft errors.

\end{abstract}


\section{Introduction}

\IEEEPARstart{V}{ision} Transformers (ViTs) that take advantage of the self-attention mechanism to model long-range dependencies of input images achieve impressive accuracy in main-stream computer vision tasks especially on large datasets compared to CNNs, and attract considerable attentions of researchers from both industry and academia recently. Motivated by the success of ViT \cite{dosovitskiy2020image}, numerous variants have been proposed to improve ViTs from different angles such as novel training strategies \cite{touvron2021training} \cite{tang2022self}, effective self-attention mechanisms \cite{wu2022pale} \cite{guo2022beyond}, and combining with classical neural network layers\cite{wu2021cvt, maaz2023edgenext, weng2022semi}. While these ViTs works mainly optimize performance and accuracy \cite{wortsman2022model} \cite{ham20203}, reliability issues induced by soft errors remain overlooked generally. On the other hand, the occurrence of soft errors per chip increases continuously \cite{baumann2005radiation} \cite{borkar2005designing} due to the increasing transistor density and shrinking feature sizes of a single transistor, which makes soft errors almost inevitable for large-scale VLSI designs supporting ViTs. Hence, reliability of ViTs in presence of soft errors needs to be explored especially for safety-critical applications like autonomous driving, where reliability can even hinder the adoption of the ViTs in practice. 

The reliability of CNNs has been explored comprehensively in prior works\cite{xu2021r2f, xu2021reliability, liu2022special}, there is still a lack of reliability analysis of ViTs that are constructed with Transformer blocks and differ substantially from CNNs. Specifically, ViTs  \cite{dosovitskiy2020image} \cite{liu2021swin} with the self-attention mechanism usually incorporate global information of inputs, while CNNs\cite{he2016deep} mostly extract local features with sliding window computing over inputs. Accordingly, soft errors on inputs of ViTs become less significant and pose less influence on the model outputs. In addition, ViTs involve more complex non-linear functions like GELU and softmax in each layer, while CNNs usually have simpler non-linear functions in a layer. Basically, non-linear functions of ViTs take up much higher proportion of the entire model computing. As a result, soft errors induced computing variations in non-linear functions become non-negligible in ViTs and comprehensive reliability analysis of ViTs remains highly demanded. 

\IEEEpubidadjcol
To understand the influence of soft errors on ViTs, we explore the vulnerability of ViTs in different granularities including models, layers, modules, and patches. The vulnerability factor for the components of ViTs is defined to the model accuracy increase between accuracy of model without protection and accuracy of model with the component under evaluation perfectly protected. It also reveals the model accuracy penalty if the target component suffers soft errors. While ViTs involve much more non-linear functions that are not considered in prior fault injection frameworks \cite{reagen2018ares} \cite{mahmoud2020pytorchfi} targeting on CNNs with much less and simpler non-linear functions, we utilize a unique operation-wise fault injection framework\cite{xue2022winograd} rather than these neuron-wise fault injection frameworks to obtain the vulnerability of all the different components of ViTs.

Based on the vulnerability analysis of ViTs, we further propose corresponding fault-tolerant approaches to protect the linear functions including GEMM and FC, and non-linear functions including GELU and softmax respectively. For the linear functions of ViTs that are generally implemented with GEMM, we propose a lightweight block-wise algorithm-based fault tolerance (LB-ABFT) approach to suit the GEMM with different sizes in ViTs. For the non-linear functions, we apply a range-based protection approach to avoid acute computing results and alleviate the negative influence of soft errors. The proposed fault-tolerant approaches show significant model accuracy improvement with minor computing overhead.

The contributions can be summarized as follows:
\begin{itemize}
\item We perform a comprehensive reliability analysis of ViTs from different granularities including models, layers, modules, and patches in presence of various soft errors for the first time, which can be potentially utilized to guide fault-tolerant design of ViTs.

\item With the reliability evaluation, we compare the fault tolerance of CNNs and ViTs in detail and observe that linear functions in both ViTs and CNNs are more fragile, but the non-linear functions in ViTs pose more negative influence on the model accuracy in presence of soft errors than in CNNs. In addition, vulnerability factors across the inputs of ViTs are generally evenly distributed, which differs substantially from that of CNNs. 

\item Based on the reliability analysis, we propose a LB-ABFT and a range-based approach to mitigate soft errors in linear functions and non-linear functions of ViTs respectively. Our experiment results show that the proposed LB-ABFT approach generally enhances the model accuracy significantly and particularly reduces the computing overhead considerably when compared to the  standard ABFT implementation. The hybrid approach shows significant accuracy improvement and can almost fully recover the accuracy even when the accuracy drops by 50\%.
\end{itemize}

\section{Background and Related Work}

In this section, we provide a brief description of ViTs and review prior work about neural network vulnerability analysis and fault-tolerant neural network design against soft errors.

\subsection{Vision Transformers}
\begin{figure}[!t]
\vspace{-0.5em}
\centering
\includegraphics[width=0.42\textwidth]{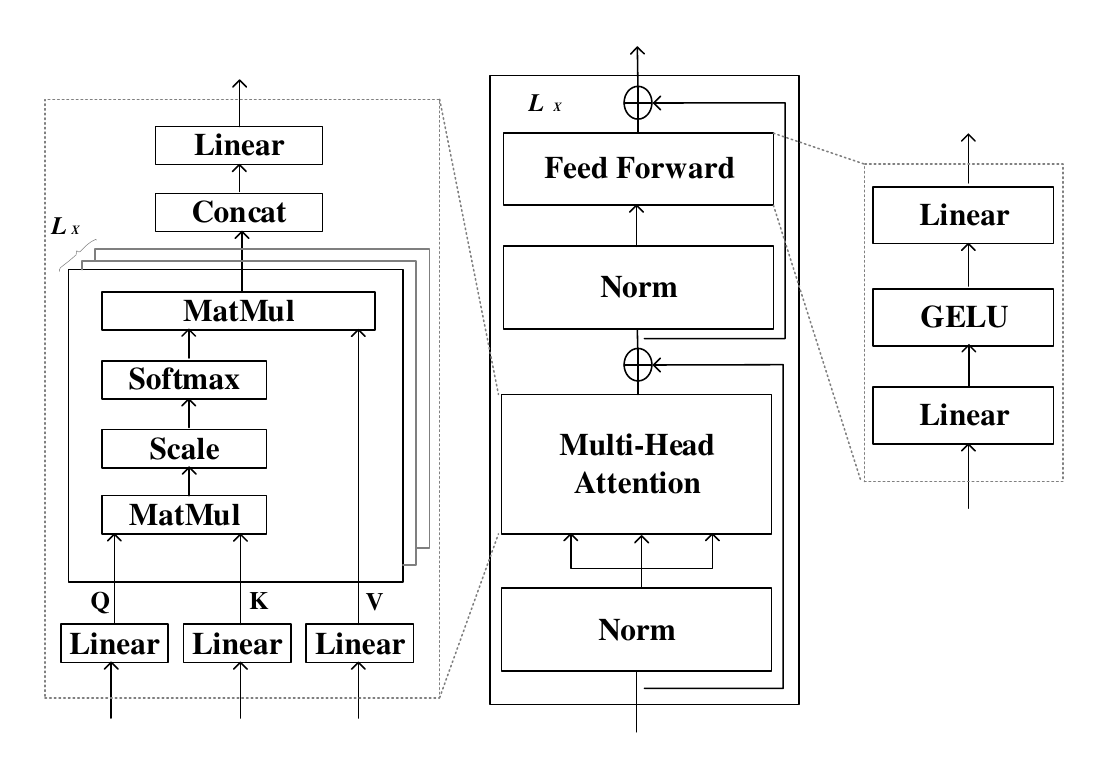}
\vspace{-1em} 
\caption{Typical Vision Transformer Architecture.}
\vspace{-1em} 
\label{fig:vit-arch}
\end{figure}

Transformer architecture that leverages self-attention mechanism to capture non-local relationships between all input sequence elements effectively achieves state-of-the-art performance and has become the de-facto standard for typical natural language processing (NLP) tasks. Inspired by the great success of the Transformer architecture in NLP, researchers also applied Transformer to computer vision tasks including classification, detection, segmentation, and demonstrated superior performance of ViTs \cite{dosovitskiy2020image}. A typical ViT architecture as presented in Figure \ref{fig:vit-arch} is composed of multiple Transformer encoders and each encoder consists of a set of connected modules including multi-head attention module (MHA), feed-forward module (FF), residual connections, and layer normalization. MHA is utilized to model long-range dependencies of inputs, mainly including GEMM, FC, softmax, and some simple layers like scale and concat. It can be formalized as Equation \ref{eq:vit} where $d_k$ stands for the dimension of query and key vectors. FF usually consists of two linear FC layers and a nonlinear activation function i.e. GELU. In addition, another major difference compared to typical CNNs is the image pre-processing layer. As for the residual connections and layer normalization, they are the same with that in typical CNNs. In addition, ViTs need to split input images into a sequence of non-overlapped patches with linear projection.   
\begin{equation} \label{eq:vit}
Attention\left(Q,K,V\right)\;=\;Softmax\left(\frac{QK^T}{\sqrt{d_k}}\right)V
\end{equation}

In the past few years, a variety of different models of ViTs have been proposed.  \cite{dosovitskiy2020image} is the pioneering work of vision Transformer and demonstrates that pure Transformer architectures are capable to outperform state-of-the-art CNN models on typical vision tasks with large datasets. Swin Transformer \cite{liu2021swin} is a hierarchical Transformer that has representation computed with shifted windows to provide a flexible modeling at various scales with linear computational complexity of image sizes. Deeper Vision Transformer (DeepViT) \cite{zhou2021deepvit} proposed to re-generate the attention maps to increase their diversity at different layers and enable deeper scaling. Class-Attention in Image Transformers (CaiT)\cite{touvron2021going} proposed to utilize class attention in the last few layers and insert CLS (classes token) as part of the class attention in the later stage of the network. DeiT \cite{touvron2021training} investigated several training strategies to learn on smaller datasets. T2T-ViT \cite{yuan2021tokens} proposed a Tokens-to-Tokens module (T2T module) to learn patch partitioning and reduce model size. Despite the tremendous efforts on ViTs, prior work mainly focus on the Transformer architecture design for the sake of higher performance and prediction accuracy, reliability of ViTs is generally overlooked.

\subsection{Reliability Analysis of Deep Learning}

Soft errors become inevitable with the growing transistor density and continuously lowering voltage of large-scale VLSI designs. Soft errors, also called transient faults, are failures caused by high-energy neutron or alpha particle strikes in integrated circuits, which may be originated from packaging materials or cosmic rays. Soft errors typically will result in a bit flip from 0 to 1 or 1 to 0, so bit-flip error is usually utilized to model soft errors in fault simulation. We will utilize soft errors or bit-flip errors interchangeably across this paper. The soft errors can propagate along with the deep learning computing data flow, spread to more operations, and generate incorrect computing results, which may induce considerable prediction accuracy loss of deep learning models that eventually rely on the silicon-based computing engines such as deep learning accelerators, GPUs, and CPUs.

Quantifying the influence of soft errors and understanding the reliability of deep learning is an essential step to protect against the soft errors especially for safety-critical applications like autonomous driving and robotics. There have been many prior works that evaluated the reliability of CNNs subjected to soft errors from different angles. For example, \cite{li2017understanding} studied the resilience of CNNs with different data types, values, data reuses, and types of layers to guide the fault tolerance design. \cite{reagen2018ares} explored the relationship between fault rate and model accuracy under various setups such as quantization, layer type, and network structure. \cite{sabbagh2019evaluating} mainly evaluated the impact of model compression especially pruning and quantization on the resilience of CNNs. The authors in \cite{xu2021reliability} and \cite{xu2020persistent} leveraged FPGA based fault injection to evaluate the influence of persistent faults on neural network models and particularly investigated the faults on both control path and data path to obtain the reliability from a system point of view. \cite{he2020fidelity} proposed a new reliability evaluation framework that brings both deep learning accelerator architectural and compilation details to the high level deep learning processing for more accurate reliability and model accuracy evaluation. Some of the reliability evaluation work further explored the reliability difference among components of deep learning models such that they can be utilized to guide selective protection with less protection overhead \cite{libano2018selective, wei2020g, mahmoud2021optimizing, ruospo2022selective}. For instance, \cite{xue2022winograd} investigated the reliability difference among the different operations and layers, \cite{mahmoud2020hardnn} explored the reliability difference among the different channels, and \cite{schorn2018accurate} explored the reliability difference among the different neurons with a classical layer-wise relevance propagation technique. These approaches generally enable more fine-grained selective protection compared to prior layer-wise protection in \cite{R2F2021TVLSI, bertoa2022fault, ping2020sern}. More reliability evaluation study of deep learning can be found in recent surveys \cite{liu2022special}\cite{mittal2020survey}.

\subsection{Fault Tolerance of Deep Learning}

To protect deep learning processing against errors in silicon, many fault-tolerant approaches have been extensively studied. Classical redundancy approaches such as dual modular redundancy (DMR) and triple modular redundancy (TMR) can potentially alleviate the influence of soft errors, but straightforward DMR or TMR require substantial computing overhead and they are usually combined with vulnerability analysis such that fragile components can be selectively protected to reduce the overhead \cite{schorn2018accurate, R2F2021TVLSI, bertoa2022fault}. Algorithm-based fault tolerance (ABFT) techniques \cite{Ozen2019sanity-check, huang1984algorithm, zhao2020ft, kosaian2021arithmetic}, originally developed for matrix-matrix multiplication based on checksum, have also been explored to detect and recover faults in deep learning without model modification, which are particularly suitable for commercial-off-the-shelf computing engines. Different from the above approaches that usually keep deep learning models intact, many prior approaches explored the inherent fault tolerance of deep learning models with distinct strategies such as model architecture design  \cite{ning2021ftt}, model parameter retraining \cite{dutta2019codenet, hacene2019training, ghavami2022fitact}, and quantization restriction \cite{chen2021low, gambardella2019efficient, zhan2021improving}.

In summary, we notice that deep learning models must be aware of the errors in silicon to ensure resilient and safety processing, and reliability evaluation as well as fault-tolerant design of neural networks have been intensively explored. However, it will be inefficient if these works are applied to ViTs directly. For instance, the completely different transformer blocks in ViTs compared to CNNs can probably affect the fault tolerance of models. Thereby, rather than reusing prior fault-tolerant approaches developed for CNNs directly, we need a comprehensive investigation of the fault tolerance of ViTs and decide the appropriate fault-tolerant approaches for ViTs. At the same time, prior ViTs works mainly focused on design for the performance and prediction accuracy, the reliability of ViTs remains generally overlooked, which can be a key barrier that hinders the adoption of ViTs on reliability-sensitive applications.

\begin{figure*}[!t]
\vspace{-0.5em}
\centering
\includegraphics[width=0.9\textwidth]{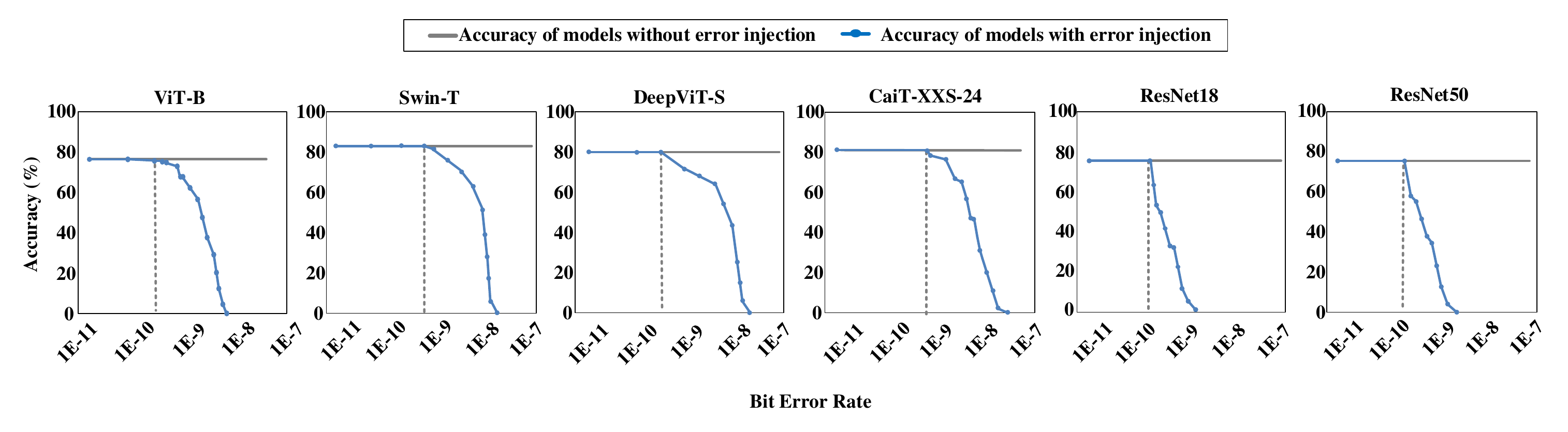}
\vspace{-1em}
\caption{Top-1 model accuracy under different BER setups.}
\vspace{-0.5em}
\label{fig:network-acc}
\end{figure*}

\begin{figure*}[!t]

\centering
\includegraphics[width=0.9\textwidth]{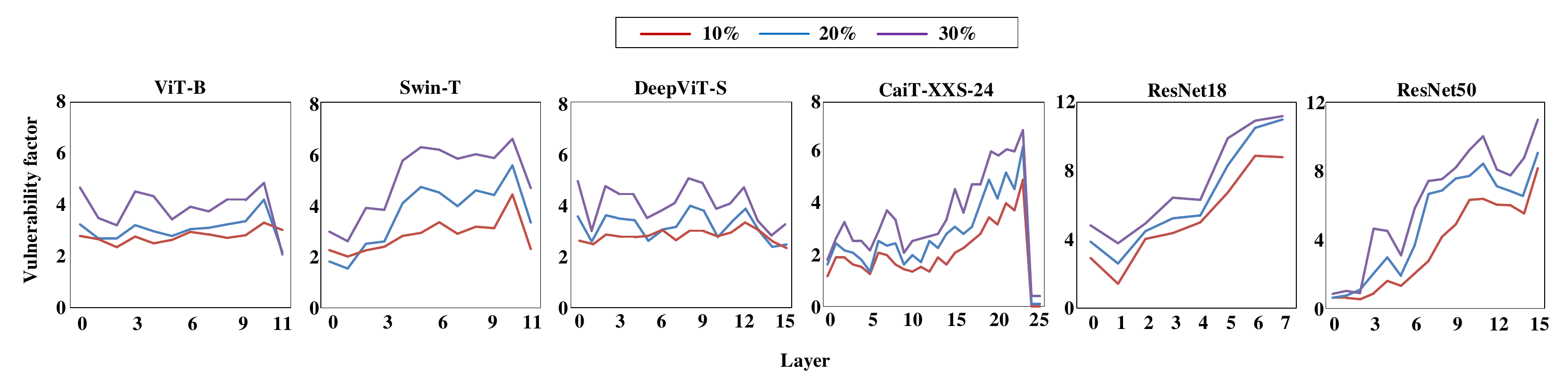}
\vspace{-1em}
\caption{Layer-wise vulnerability factors.}
\vspace{-1em}
\label{fig:layer-reliability}
\end{figure*}

\section{ViTs Reliability Evaluation} \label{sec:reliability-evaluation}

We mainly quantify the reliability of ViTs from the perspective of network architectures and investigate the vulnerability of ViTs from different granularities including models, layers, modules, and patches respectively.

\subsection{Evaluation Setups}
The major experiment setups include datasets, ViTs models benchmark, fault models, and fault injection framework.

\textbf{Dataset.} We take the well-known ImageNet-1K as the dataset \cite{deng2009imagenet} and randomly select 10000 images for the accuracy evaluation. The resolution of each image in the dataset is $256 \times 256$.

\begin{table}[htb]
\caption{Major Parameters of the ViTs benchmark\label{tab:table1}}
\centering
\begin{minipage}{\textwidth}
\begin{tabular}{c c c c c}
\hline
Network & Layer Num & Head Num & eDim\footnote{eDim refers to embedding dimension.} & Patch Size \\
\hline
ViT-B & 12 & 12 & 768 & $16\times16$\\
Swin-T & [2, 2, 6, 2] & [3, 6, 12, 24] & 96 & $4\times4$\\
DeepViT-S & 16 & 12 & 396 & $16\times16$\\
CaiT-XXS-24 & 24+2 & 4 & 192 & $16\times16$\\
\hline
\end{tabular}
\end{minipage}
\end{table}

\textbf{ViTs Models.} We select four representative ViTs models as the benchmark for reliability evaluation, and select ResNet18 and ResNet50\cite{he2016deep} for comparison. The detailed configurations of these models such as the number of layers, number of heads, embedding dimension, and patches are summarized in Table \ref{tab:table1}. ViTs include ViT-B\cite{dosovitskiy2020image}, Swin-T\cite{liu2021swin}, DeepViT-S\cite{zhou2021deepvit}, and CaiT-XXS-24\cite{touvron2021going}. All the neural network models are quantized with int8. Top-1 accuracy is utilized as the evaluation metric.

\textbf{Fault Models.}  We utilize bit flip model to characterize typical soft errors. Particularly, following prior reliability analysis works \cite{reagen2018ares}\cite{xue2022winograd}\cite{li2017understanding} , we also use bit error rate (BER), which represents the ratio of bit flip errors over the total number of bits of operations in the model as the soft error intensity metric. It ranges from 1E-11 to 1E-7 which can cover the major model accuracy degradation curve.  

\textbf{Error Injection.} We adopt an operation-wise error injection method proposed in \cite{xue2022winograd} for the reliability evaluation. It is essentially an approximate implementation of bit-flip or soft errors. We have it verified with both prior neuron-wise fault injection frameworks and a fault-injection framework with architecture details \cite{xue2023winograd}. It is built on top of PyTorch and injects random bit flip errors to outputs of the major arithmetic operations i.e. multiplication and addition in each model, which focuses on the computing of the models rather than a specific computing engine. It covers both linear and non-linear functions in the models.

\textbf{Vulnerability Factor.} To differ the reliability of different ViTs components, we utilize vulnerability factor as the metric, which refers to the model accuracy difference between model with only the target component perfectly protected and the model without any protection.

\textbf{Hardware Platforms.} All the evaluation experiments are performed on a server equipped with two 24-core@2.5GHz Intel Xeon processors, 512GB memory, and four PH402 SKU 200 GPU cards.

\subsection{Model-wise Reliability Evaluation}

In this sub section, we evaluate the Top-1 accuracy of benchmark models under different BER setups ranging from 1E-11 to 1E-7, and utilize the accuracy degradation curves to characterize the resilience of the models subjected to bit flip errors. The results is shown in Figure \ref{fig:network-acc}. It reveals that the general trend of the accuracy curves of different models are similar. Basically, most of the computing errors can be tolerated and the model accuracy generally remains steady or drops slightly when the BER is low. When the BER further increases, the model accuracy drops sharply as the distributed errors reach certain threshold and corrupt the models. In addition, although the general trend of the accuracy curves of different models are similar, model architecture poses substantial influence and results in different accuracy degradation. 

Specifically, we observe that the model accuracy of enhanced ViTs models starts to drop at higher BER than ResNet. For instance, ResNet18 starts to drop when the BER is at 3E-10, while ViTs models such as Swin-T and CaiT-XXS-24 start to drop when the BER is at round 9E-10. In addition, the accuracy degradation curve of ResNet18 and ResNet50 are sharp and the accuracy drops to zero very fast. While the accuracy degradation curves of ViTs models are relatively more smooth in general, which demonstrates that ViTs models are more resilient when subjected to the same bit error rate. Nevertheless, due to the much higher computing overhead, straightforward redundancy based protection can incur considerable overhead. Hence, more fine-grained fault analysis is required to understand the fault tolerance of ViTs comprehensively.

\begin{figure*}[!t]
\vspace{-0.5em}
\centering
\includegraphics[width=0.9\textwidth]{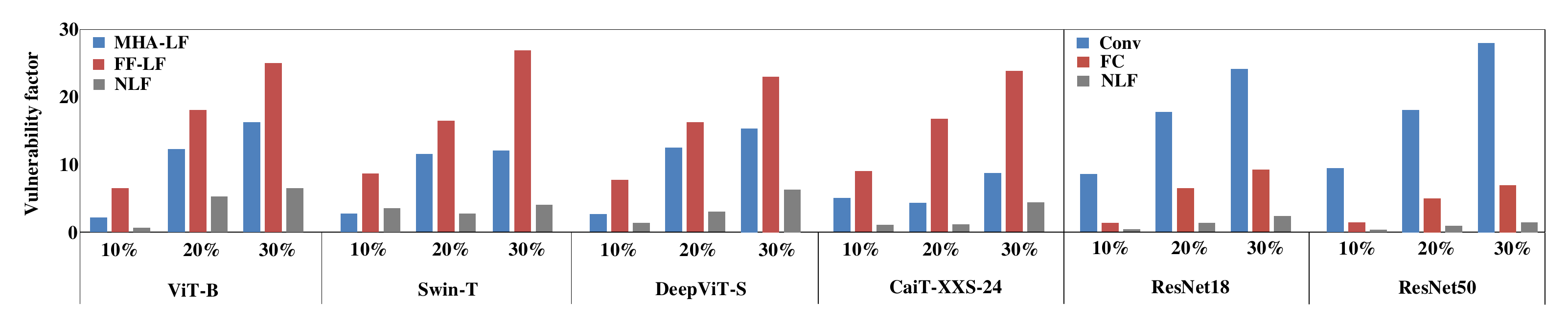}
\vspace{-1em}
\caption{Module-wise vulnerability factors.}
\vspace{-0.5em}
\label{fig:module-wise}
\end{figure*}

\subsection{Layer-wise Reliability Evaluation}

To gain insight of the reliability of ViTs, we take Transformer blocks as layers, and investigate layer-wise vulnerability factors in this sub section. Specifically, we utilize the model accuracy difference between model without any protection and the same model that only has the target layer set to be fault-free as the metric to quantify the vulnerability of the layer. 
Larger vulnerability factors indicate that the corresponding layers are more sensitive to the bit-flip errors and need to be protected with higher priority in general. Since vulnerability factor is closely related with BER, we set BER to be the value that leads to moderate accuracy loss i.e. 10\%, 20\% and 30\% accuracy loss, which can be potentially addressed with reasonable overhead. The layer-wise vulnerability factor of the different ViTs models is shown in Figure \ref{fig:layer-reliability}. It reveals that the variation of the layer-wise vulnerability factors vary across the layers. However, the vulnerability factor variation across different layers of ViTs models are lower than that of ResNet, which is probably caused by the global feature extraction based on the self-attention mechanism in ViTs and insignificant FLOPs difference between different layers.

\begin{figure*}[!t]
\centering
\includegraphics[width=0.9\textwidth]{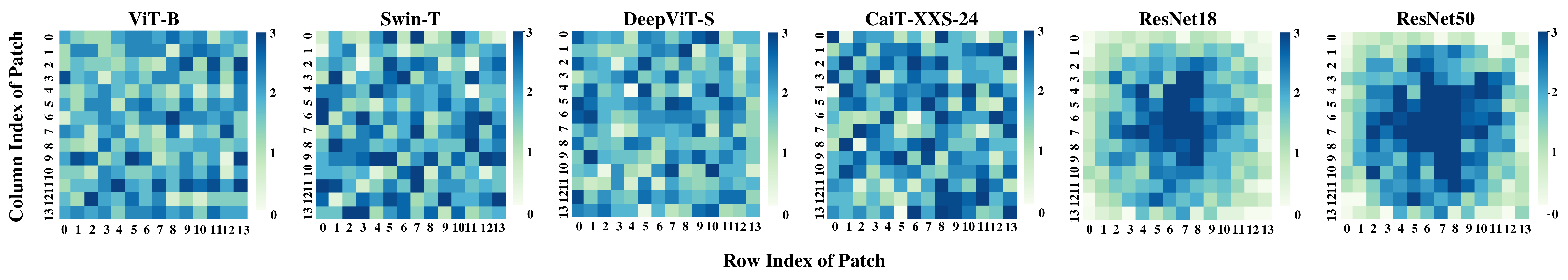}
\vspace{-1em}
\caption{Heatmap of patch-wise vulnerability factors.}
\vspace{-1em}
\label{fig:patch-wise}
\end{figure*}

\subsection{Module-wise Reliability Evaluation}
Different from the layer-wise reliability analysis, we also explore the reliability of ViTs models from the perspective of operation modules. ViTs can be roughly divided into linear functions such as GEMM and FC, and non-linear functions (NLF) such as softmax and GELU. Particularly, we denote linear functions in MHA and FF as MHA-LF and FF-LF, respectively. In contrast, ResNet18 and ResNet50 are divided into convolution modules (Conv), FC modules and NLF modules like ReLU. For the error injection to non-linear functions, we inject random bit-flip errors to the basic multiplication and addition operations. Similar to prior layer-wise vulnerability factors, we also utilize the vulnerability factors to characterize the sensitivity of the different modules to the bit flip errors. As for the BER used in the evaluation, we set BER to be the same as that used in layer reliability analysis. 

The experiment result is shown in Figure \ref{fig:module-wise}. It can be seen that the module-wise vulnerability factors vary substantially across the different modules. Particularly, FF with a large number of operations in ViTs model turns out to be the most fragile part, which indicates that FF is generally more sensitive to bit-flip errors. In contrast, MHA generally has much lower vulnerability factor than FF, despite the non-trivial amount of operations. This is mainly because that the softmax limits the output value within the range  between 0 to 1 and suppresses the influence of errors substantially. For ResNet, the vulnerability factors of Conv are significantly higher as expected because they take up the majority of operations in ResNet. In addition, NLF that takes up only a fraction of the overall operations of ViTs, but it still shows some vulnerability in ViTs models. As for ResNet, the proportion of NLF is much smaller, exhibiting a relatively low vulnerability factor.

\subsection{Patch-wise Reliability Evaluation}

To study the influence of soft errors on input images, we follow the ViTs patch setups and investigate the patch-wise vulnerability factors of the ViTs models. To compare with convolutional neural network models, we also analyze the inputs of ResNet18 and ResNet50 with the granularity of patches and the patch is set to be $16\times16$. Similar to the layer-wise vulnerability factor, we take the accuracy difference between models with the target input patch perfectly protected and models without input protection as the patch-wise vulnerability factor. Since the vulnerability is closely related with the error rate, we set BER to be the value that leads to 10\% accuracy loss. The patch-wise vulnerability factor is presented with heatmap as shown in Figure \ref{fig:patch-wise}. It shows that the distribution of the vulnerability factors of ViTs and ResNet are quite different. Since objects in ImageNet are usually located in the center of input images, errors on pixels located in the center pose more negative influence on ResNet model accuracy as expected. The patch vulnerability factors of ViTs show a relatively even distribution and vulnerable patches can scatter across the input images. This is because that ViTs leverage the self-attention mechanism to obtain global information rather than local features.

\section{Fault-Tolerant Approaches for ViTs}

According to the analysis in Section \ref{sec:reliability-evaluation}, we notice that the major reliability challenges of ViTs models for linear operations and non-linear operations differ substantially. For linear operations that are  typically implemented with GEMM take up considerable computing overhead, so they need to be protected with a lightweight approach. NLF involves much less computing but still has influence on the model reliability, so NLF protection needs to emphasize more on the accuracy rather than computing overhead. In this work, we propose an adaptive yet lightweight block-wise ABFT (LB-ABFT) approach to protect the linear operations in ViTs and we adopt a range-based approach to explicitly suppress the outputs of NLFs that can vary considerably in presence of bit-flip errors. Details about the fault-tolerant approaches will be illustrated as follows.

\subsection{Fault-tolerant Approach for Linear Operations}

\begin{figure*}[!t]
\vspace{-1.5em}
\centering
\includegraphics[width=0.9\textwidth]{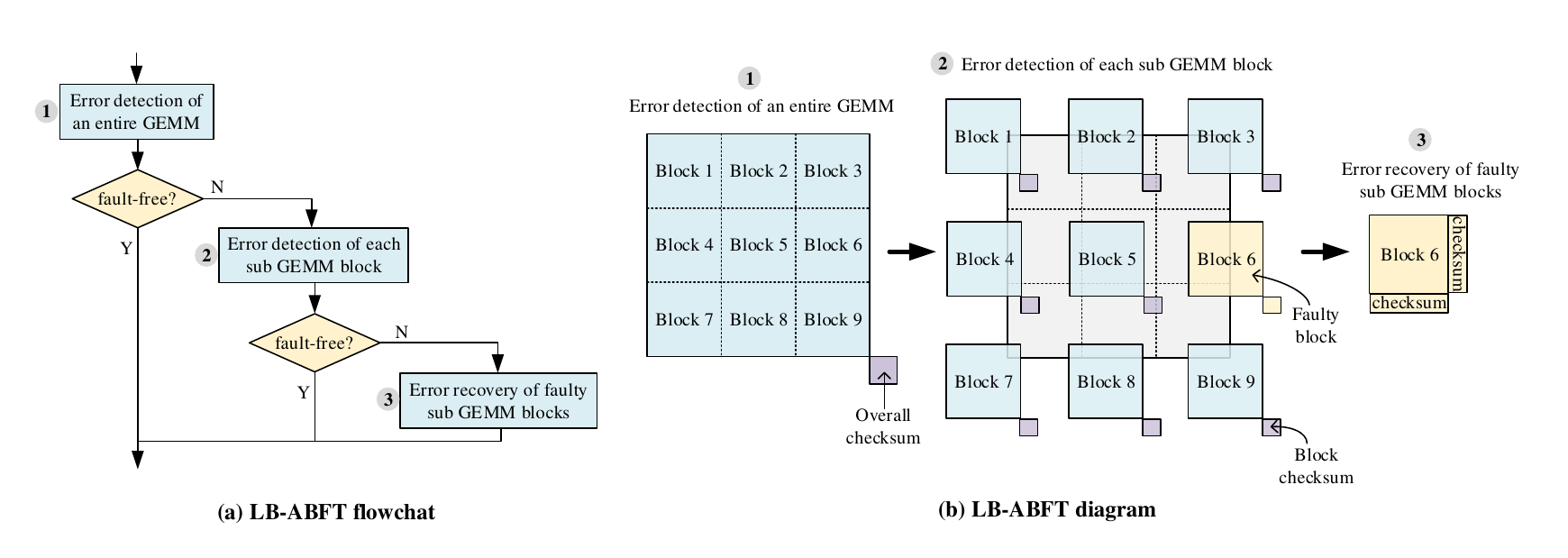}
\vspace{-1em}
\caption{(a) LB-ABFT flowchat. (b) LB-ABFT diagram.}
\vspace{-1em}
\label{fig:LBABFT}
\end{figure*}

Standard ABFT algorithm essentially leverages a checksum mechanism to obtain sum of outputs with different approaches such that they can be utilized for both the fault detection and recovery. Standard ABFT can be divided into an error detection stage and an error recovery stage. Take an $n \times n$ GEMM as an example, error recovery requires $2n^2$ MACs and it is more expensive compared to error detection that requires only $n$ MACs. Error recovery is invoked only when computing errors are detected, but it is beneficial only when there is only one output error in each row or column due to the limited error correction capability. Thereby, it is inefficient to adopt standard ABFT when multiple computing errors occur in a single row or column at higher BER. Additionally, we observe that the sizes of GEMM in ViTs models can vary dramatically, and direct implementation of standard ABFT without being aware of the matrix sizes or the error rate can induce frequent error recovery procedures and low reliability. Meanwhile, layer vulnerability factors can vary, which further complicates the protection. 

\algdef{SE}[DOWHILE]{Do}{doWhile}{\algorithmicdo}[1]{\algorithmicwhile\ #1}%
\begin{algorithm}[!ht]
\footnotesize
\caption{LB-ABFT Algorithm}
\hspace*{0.02in} {\bf Input:} A model with $L$ layers and each layer involves $N$ matrix multiplications, the target model accuracy is set to be $ACC$ under bit error rate $r$.\\
\hspace*{0.02in} {\bf Output:} Determine the LB-ABFT protection strategy that minimizes the LB-ABFT computing overhead and fulfills the specified model accuracy $ACC$ as well as computing overhead limit $C_0$.
\begin{algorithmic}[1]

\State Optimize LB-ABFT protection operation sequence $S$ ← $\varnothing$, where $(m_{i,j}, p_{i,j}) \in S$, $m_{i,j}$ refers to the $j$th GEMM in $i$th layer of the model and $p_{i,j}$ stands for the partition setups of the GEMM. Note that $i \in (1,...,L)$, and $j \in (1,...,N)$. 

\State Obtain the vulnerability factors $LV$ of all the layers.
\State Sort $LV$ with descending order and obtain the protection order $LR$ of the layers.
\State $i$ ← $1$, $C$ ← $0$
\While{($acc < ACC$ and $C < C_0$)}

      \State $LR_{i}$ refers to the layer ID that ranks at $i$th in vulnerability factor.

      \State obtain the vulnerability factors $MV$ of all the matrix in $LR_{i}$th layer.

      \State Sort $MV$ with descending order and obtain the protection order $MR$ of all the GEMMs in this layer.
      
      \For {$j$ = 1 to $N$}
            \State $p_{LR_{i},MR_{j}}, c$ ← $OptBlockSizeSelection$($m_{LR_{i}, MR_{j}}$, $r$).
            
            \State Apply LB-ABFT based on $p_{LR_{i},MR_{j}}$ to the GEMM $m_{LR_{i},MR_{j}}$.
            
            \State Append $p_{LR_{i},MR_{j}}$ to $S$.          
            \State Obtain model accuracy $acc$.
            \State Computing overhead $C$ ← $C$ + $c$
            \If {($acc > ACC$ or $C > C_0$)}
            \State break;
            \EndIf
      \EndFor      
      \State $i$ ← $i+1$      
\EndWhile
\State \Return LB-ABFT protection sequence $S$, model accuracy $acc$, and computing overhead $C$.

\end{algorithmic}  
\end{algorithm}

\algdef{SE}[DOWHILE]{Do}{doWhile}{\algorithmicdo}[1]{\algorithmicwhile\ #1}%
\begin{algorithm}[!ht]
\footnotesize
\caption{OptBlockSizeSelection}
        \hspace*{0.02in} {\bf Input:} The dimensions of the two input matrices are $m \times n$ and $n \times p$ respectively and the bit error rate is $r$\\
\hspace*{0.02in} {\bf Output:} Optimized block size $b$ and computing overhead $c$.
\begin{algorithmic}[1]
\State $B$ stands for all the possible block size setups.
\State $C$ stands for the computing overhead with different block sizes.
\For {each possible block size setup in B}
      \State $C[i]$ ← \Call{LB-ABFTOverhead}{$B[i]$, $r$}
\EndFor 
\State Find optimized block size $b$ that leads to minimum computing overhead $c$ in C.

\State \Return $b$, $c$      
\State 
\Function {LB-ABFTOverhead}{$b$, $r$} 

\State Number of sub matrices $blk$ ← $\lceil \frac {m}{b} \rceil$ $\times$ $\lceil \frac {n}{b} \rceil$ $\times$ $\lceil \frac {p}{b} \rceil$
\State GEMM error probability $EP_{M}$ ← 1 - $(1-r)^{m\times n \times p}$
\State Sub GEMM error probability $EP_{SM}$ ← 1 - $(1-r)^{\lceil \frac {m}{b} \rceil \times \lceil \frac {n}{b} \rceil \times \lceil \frac {p}{b} \rceil}$
\State GEMM error detection overhead $M_{detection}$ ← $n$
\State Sub GEMM error detection overhead $SM_{detection}$ ← $b$
\State Sub GEMM error recovery overhead $SM_{recovery}$ ← $2\times b \times b$
\State Total error detection overhead $T_{detection}$ ← $M_{detection}$ + $EP_{M} \times blk \times SM_{detection}$
\State Total error recovery overhead $T_{recovery}$ ← $EP_{M} \times blk \times EP_{SM} \times SM_{recovery}$

\State \Return total computing overhead $T_{detection} + T_{recovery}$
\EndFunction

\end{algorithmic}  
\end{algorithm}

To address the problem, we propose a LB-ABFT approach as shown in Algorithm 1, which selects the most vulnerable layer based on layer vulnerability factors (lines 1-3), and then prioritizes the most vulnerable GEMM in this layer for protection (lines 6-8). For the specified GEMM, we adopt a LB-ABFT strategy and divide a large GEMM into multiple sub GEMMs. As shown in Figure \ref{fig:LBABFT}, we show the flow and diagram of LB-ABFT, which consists of three processing stages. It performs the error detection of the entire GEMM first. If an error is detected, it further invokes the block-wise error detection such that faulty blocks can be determined. Finally, block-wise error recovery is performed on only faulty blocks. Among, we use the analytical algorithm $OptBlockSizeSelection$ in Algorithm 2 to adaptively select the most suitable block size for GEMM of different sizes under different BER setups and different models, instead of choosing a fixed block size to split all GEMMs (line 10). Note that the optimized block size setup is determined offline in advance. When the optimized block size is determined, we can apply the LB-ABFT strategy on the GEMM and estimate if the accuracy and computing overhead meet user requirements (lines 11-17). By iterating the above processing steps as shown in Algorithm 1, we can achieve the specified model accuracy with minimum computing overhead. When the overall computing overhead exceeds the computing overhead limit, we optimize the model accuracy under the computing overhead limit. The accuracy optimization is similar to the above computing overhead optimization procedure. At the end of this algorithm, we can obtain optimized LB-ABFT strategies for ViTs under various BER setups.

For the analytical algorithm $OptBlockSizeSelection$, we have it detailed in Algorithm 2. Basically, we utilize LB-ABFTOVERHEAD to estimate the LB-ABFT computing overhead (lines 9-19), and utilize the computing overhead as a guide to optimize the block size for each specific GEMM. Since addition and multiplication generally dominate the ABFT computing overhead of GEMMs, we utilize the number of multiplication operations and addition operations induced by ABFT as the computing overhead directly. The overhead can also be weighted depending on the target computing engines. The total computing overhead of LB-ABFT consists of error detection overhead and error recovery overhead in general. Error detection further includes detection of the entire GEMM and detection of all the sub GEMM blocks when the overall GEMM is faulty. Error recovery is the accumulated overhead of the error recovery of all the faulty sub GEMM blocks. Larger GEMM blocks usually require less error detection overhead but can incur more recovery overhead. In contrast, smaller GEMM blocks pose more detection overhead but less recovery overhead. Hence, optimized blocking setup varies with the bit error rate that determines the activation probability of the error recovery procedures and the total error recovery overhead. Since the probability of the error recovery will also be affected by the actual error distribution across the GEMM, we assume an even error distribution and take the expected computing overhead induced by both error detection and error recovery into consideration for the LB-ABFT blocking optimization metric. Essentially, the blocking algorithm is built on top of an analytical model that estimates the LB-ABFT detection overhead and recovery overhead respectively, and seeks to minimize the total computing overhead with the basic brute-force search.

\begin{figure}[!t]
\vspace{-0.5em}
\centering
\includegraphics[width=0.48\textwidth]{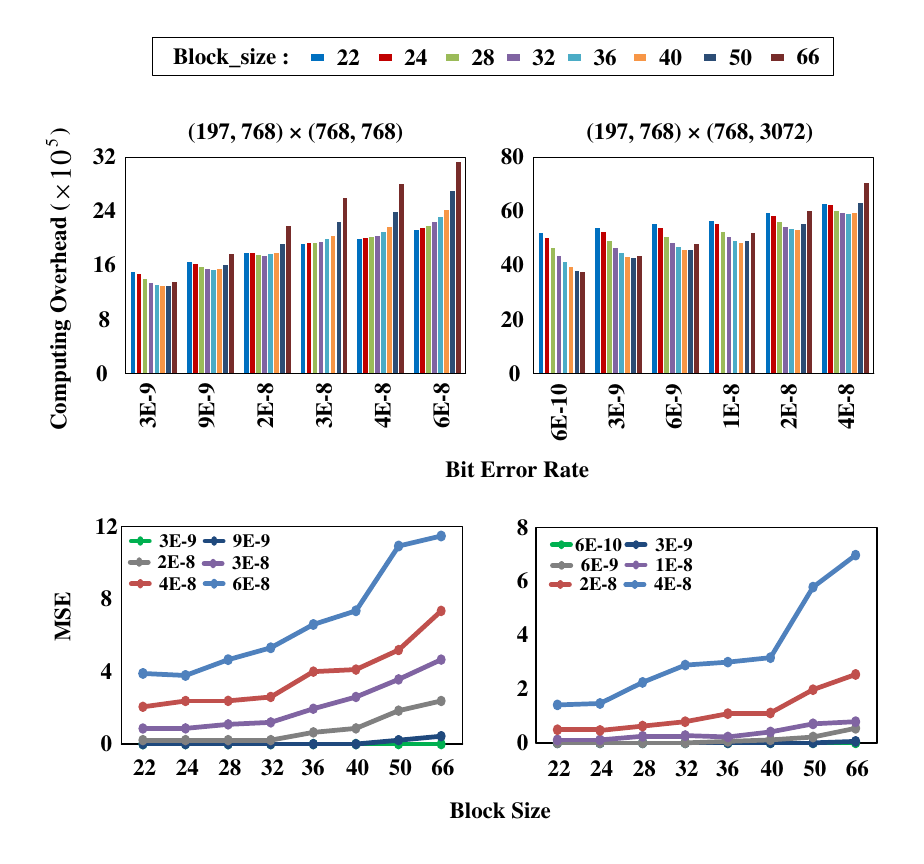}
\vspace{-1em}
\caption{LB-ABFT computing overhead and MSE under different block size setups.}
\vspace{-1em}
\label{fig:block_size}
\end{figure}

In order to verify the influence of block size on the computing overhead, we take two typical GEMMs from ViT-B as an example and evaluate the computing overhead using the proposed LB-ABFT strategy under various bit error rate setups. The experiment result as presented in Figure \ref{fig:block_size} confirms that the optimized block size cannot neither be too big nor too small as expected. In most cases, we notice that larger block size setups can result in larger computing overhead because of the frequent error recovery procedures. When the block size is smaller especially when the error rate is relatively lower, it also induces high computing overhead due to the increased error detection. At the same time, we also observe that optimized block size varies substantially over the different error rate setups. On top of the computing overhead, we utilize MSE between GEMM with error injection and GEMM without error injection as a typical computing accuracy metric and investigate the influence of block size setups on the computing accuracy. It can be seen that MSE increases relatively slow when the block size ranges from 22 to 40. Basically, these block size setups will not cause much computing accuracy difference of the GEMM and computing overhead turns out be more important for the block size selection in these cases. When the block size gets higher, MSE increases rapidly and may affect the resulting model accuracy. Hence, MSE may also affect the block size selection. Fortunately, the proposed block size optimization strategy based on computing overhead minimization generally produces moderate block size setups and shows competitive accuracy.

\begin{figure}[!t]
\centering
\includegraphics[width=0.48\textwidth]{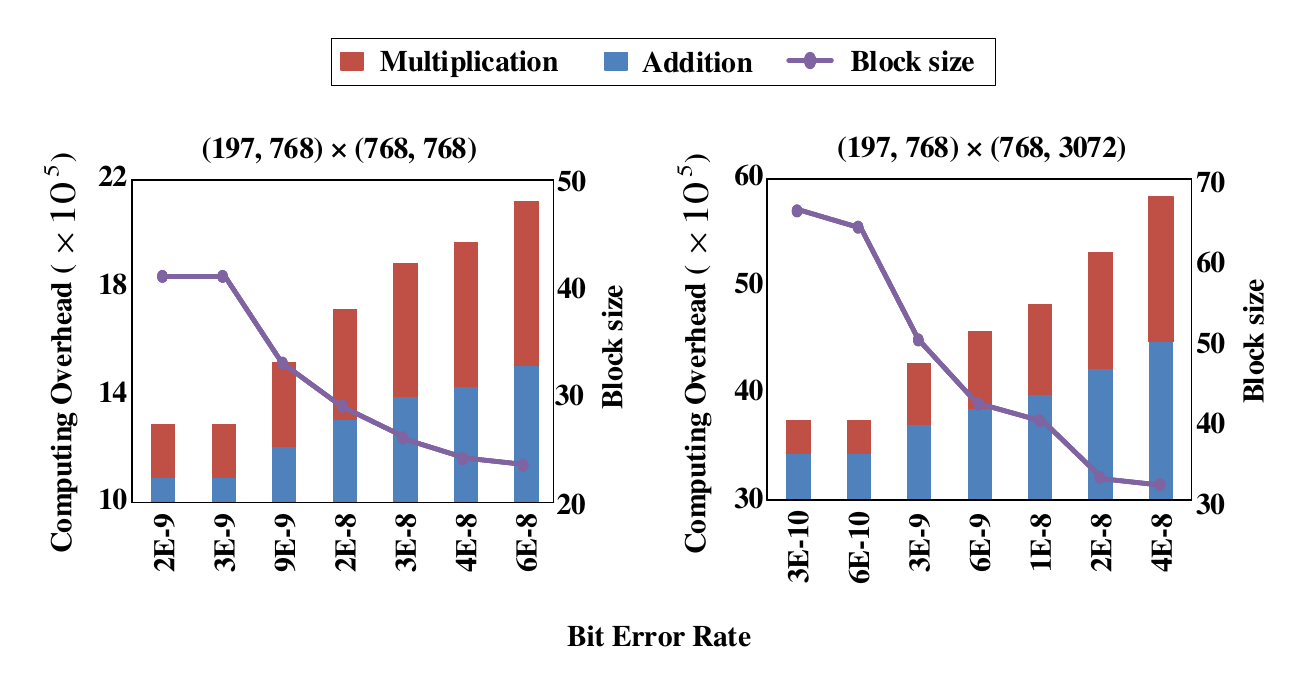}
\vspace{-1em}
\caption{Optimized block size selection analysis under different BER setups.}
\vspace{-1em}
\label{fig:gemm_mse_overhead}
\end{figure}

Based on the proposed block size optimization algorithm, we further analyze the optimized block size selection at different error rate. As shown in Figure \ref{fig:gemm_mse_overhead}, it can be seen that optimized block size is generally smaller at higher BER setups when error recovery dominates the total computing overhead. In addition, we notice that LB-ABFT computing overhead is generally higher at higher BER despite the block size optimization. The percentage of these operations depends on the size of GEMMs. It can be seen that the percentage of multiplication over the total LB-ABFT computing overhead is much higher for GEMMs with shorter second matrix.  

\subsection{Fault-tolerant Approach for Non-Linear Operations}

For NLFs in ViTs, we adopt a range-based approach \cite{ghavami2022fitact} \cite{chen2021low}  to suppress the soft errors induced computing errors in general. Specifically, for NLFs with determined range such as softmax, outputs that exceed the range between 0 and 1 will be set to be zero. For the NLFs without fixed range such as GELU, we set the range to be $((1+\alpha) \times min, (1-\alpha) \times max)$ where $min$ and $max$ represent the minimum and maximum values of the function outputs respectively and they can be obtained through statistic with a set of sampling inputs. $\alpha$ is utilized to filter out large outputs with very limited number and it is set to be 0.02 according to \cite{chen2021low}. Similarly, computing results that are out of the range will be set to be zero. Since the range-based approach only needs comparison operations and its computing overhead is trivial compared to ViTs processing.   

\begin{figure*}[!t]
\vspace{-0.5em}
\centering
\includegraphics[width=0.9\textwidth]{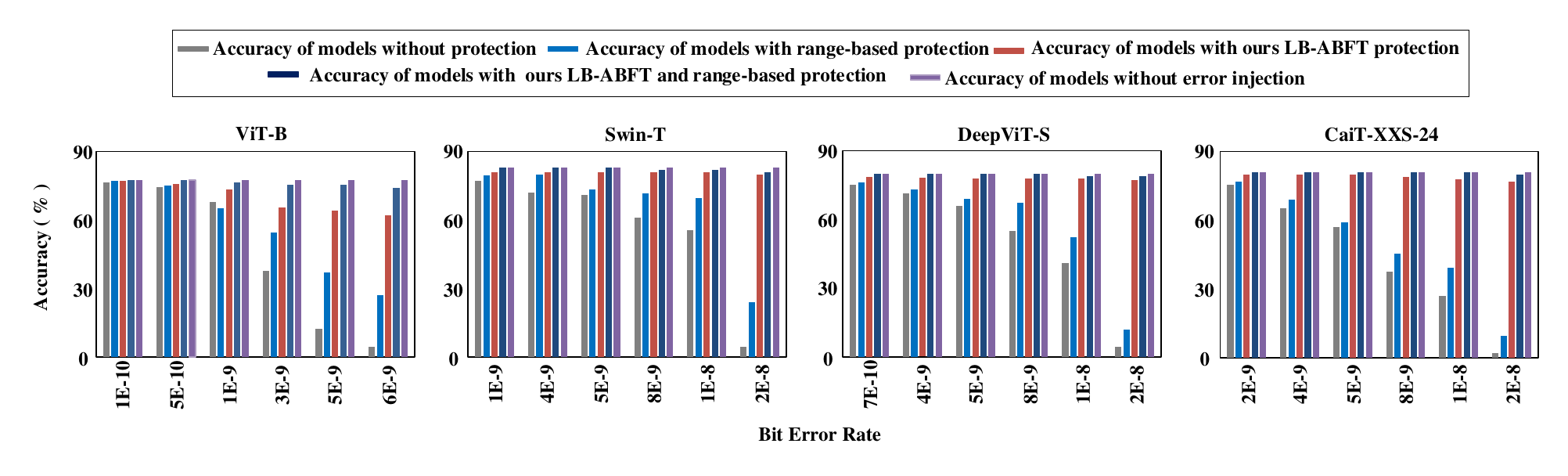}
\vspace{-1em}
\caption{Model accuracy of ViTs with LB-ABFT and range-based protection.}
\vspace{-0.5em}
\label{fig:pro_linear_nolinear_acc}
\end{figure*}

\begin{figure*}[!t]
\centering
\includegraphics[width=0.9\textwidth]{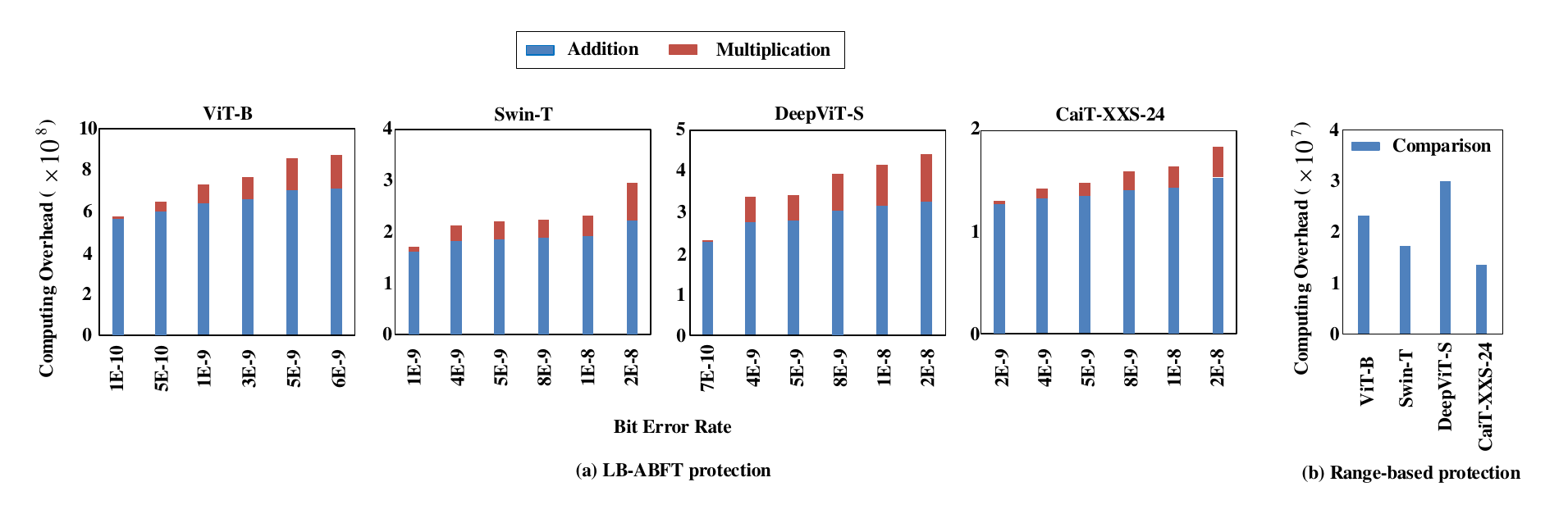}
\vspace{-1em}
\caption{Computing overhead of ViTs with LB-ABFT and range-based protection.}
\vspace{-1em}
\label{fig:pro_linear_nolinear_flops}
\end{figure*}

\subsection{Evaluation}
The proposed LB-ABFT and range-based approach are applied to protect ViTs against various soft errors. The accuracy loss relative to a clean model is set to be less than 2\% and the maximum computing overhead is set to be 2\% normalized to the total ViTs computing. To evaluate the proposed fault-tolerant approaches, we inject bit flip errors at the granularity of operations like addition and multiplication based on PyTorch such that fault injection is consistent for both linear operations and non-linear operations. It differs from prior neuron-wise fault injection frameworks \cite{reagen2018ares} \cite{mahmoud2020pytorchfi} \cite{chen2020tensorfi}  that fail to cover the non-linear functions and is open sourced on github \footnote{\url{https://github.com/xuexinghua/Operation-level-FI.git}}. With the operation-wise fault injection framework, we apply the proposed fault-tolerant approaches gradually on a set of ViTs including ViT-B, Swin-T, DeepViT-S, and CaiT-XXS-24 at different bit error rates.

We evaluate the model accuracy and compare with that of clean models and models without any protection. The comparison is shown in Figure \ref{fig:pro_linear_nolinear_acc}. It demonstrates that the proposed fault-tolerant approaches enhance model accuracy significantly in general and the accuracy can be mostly recovered even when the accuracy of the unprotected models drops by more than 50\% at higher BER setups. In addition, we notice that both linear operations and non-linear operations in ViTs can tolerate some soft errors and the model accuracy drops little when the error rate is low. However, when the error rate increases, more sharp model accuracy drop is observed if linear parts of ViTs are not protected. For instance, the model accuracy with only range-based protection gets close to zero when BER is 2E-8 for Swin-T, DeepViT-S, and CaiT-XXS-24. This is probably because that the number of primitive operations in linear functions of ViTs is much larger than that of non-linear functions in ViTs. At the same time, the influence of faults on non-linear functions remains non-trivial especially at relatively higher error rate.

\begin{figure*}[!t]
\vspace{-0.5em}
\centering
\includegraphics[width=0.9\textwidth]{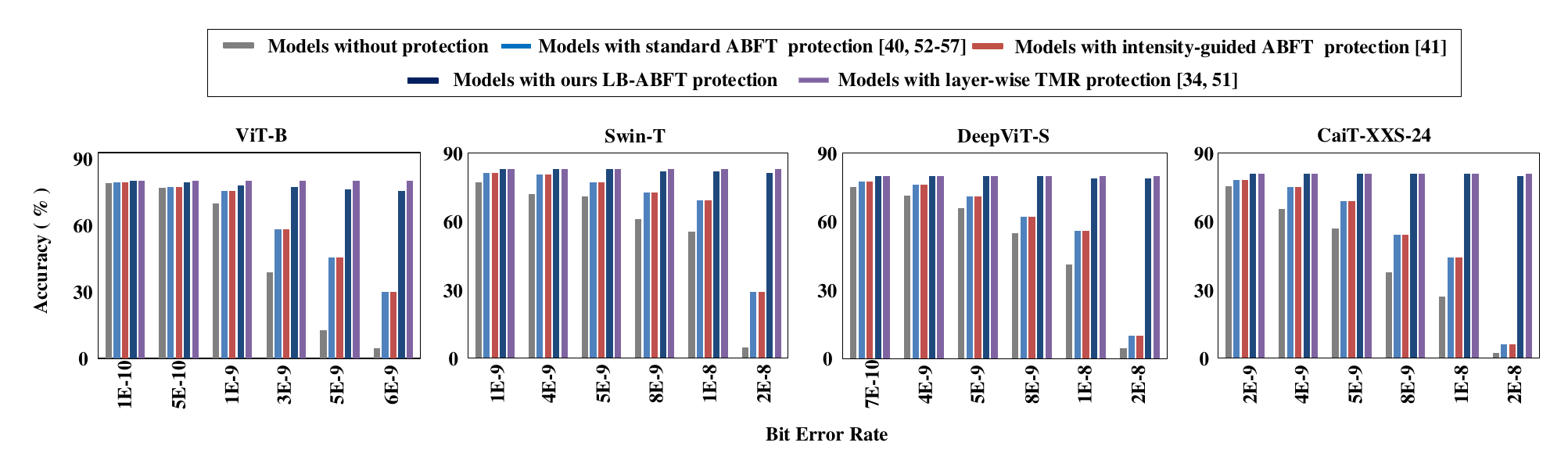}
\vspace{-1em}
\caption{Model accuracy comparison between LB-ABFT and standard ABFT \cite{zhao2020ft} \cite{safarpour2021low, roffe2020evaluation, sharif2023efficient, li2022efficient, zamani2019greenmm, liu2018fault}, intensity-guided ABFT\cite{kosaian2021arithmetic}, and layer-wise TMR \cite{R2F2021TVLSI} \cite{sabih2021fault}.}
\vspace{-0.5em}
\label{fig:pro_linear_acc}
\end{figure*}

\begin{figure*}[!t]
\centering
\includegraphics[width=0.9\textwidth]{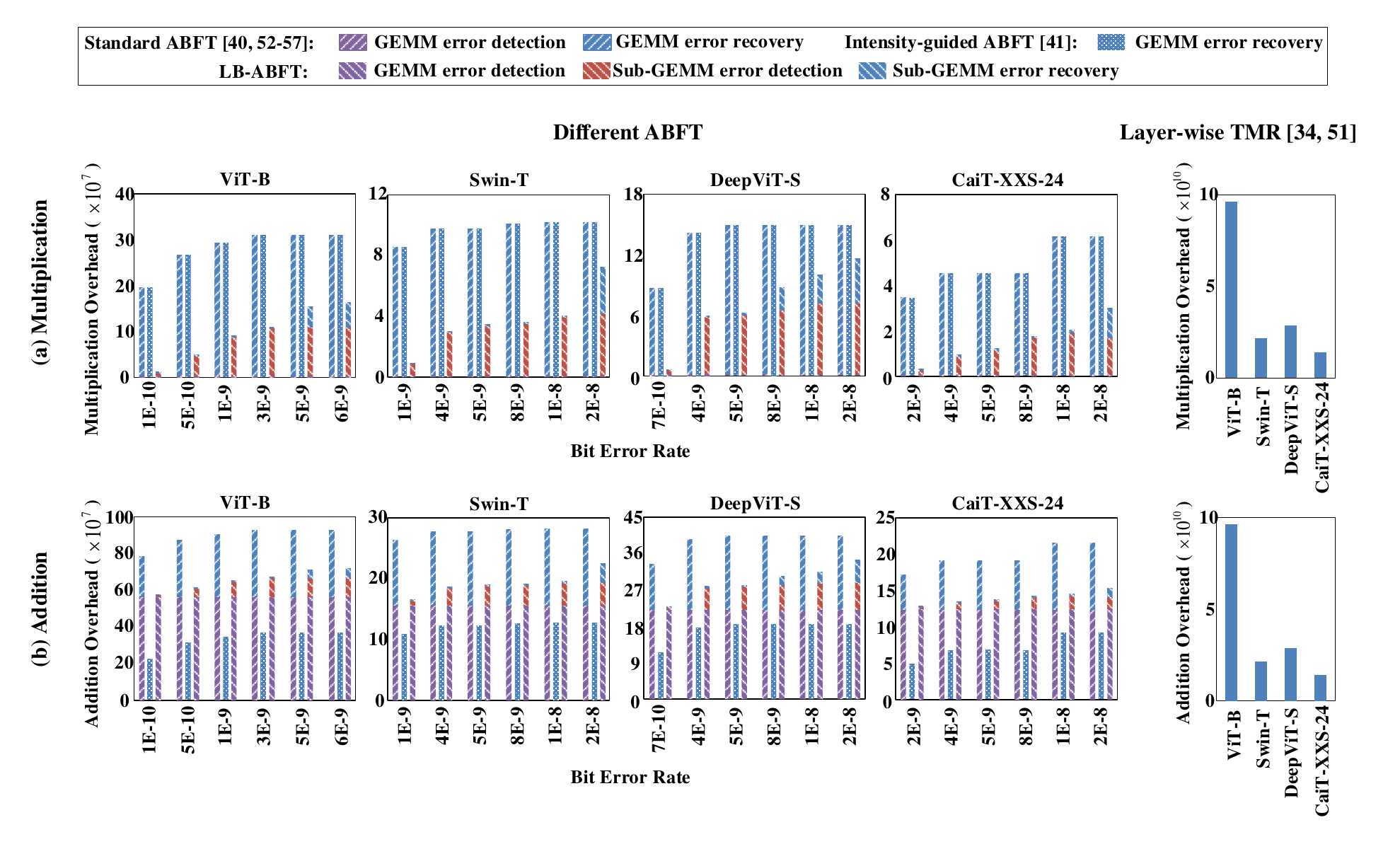}
\vspace{-1em}
\caption{Computing overhead comparison between LB-ABFT and standard ABFT \cite{zhao2020ft} \cite{safarpour2021low, roffe2020evaluation, sharif2023efficient, li2022efficient, zamani2019greenmm, liu2018fault}, intensity-guided ABFT\cite{kosaian2021arithmetic}, and layer-wise TMR \cite{R2F2021TVLSI} \cite{sabih2021fault}.}
\vspace{-1em}
\label{fig:pro_linear_flops}
\end{figure*}

Additionally, we evaluate the computing overhead of the proposed LB-ABFT and range-based protection at different BER setups in Figure \ref{fig:pro_linear_nolinear_flops}. ABFT requires multiplication and addition, while range-based approach only requires comparison. For the LB-ABFT protection, its computing overhead increases with BER. The main reason is that as the BER increases, more computing errors will occur and error recovery will be invoked more frequently. When the BER is very high, there are still multiple errors in the same sub GEMM, excessive use of LB-ABFT will not further improve the model accuracy, so we opt to relax the model accuracy requirement by using larger blocks and skipping some useless error recovery to reduce the computing overhead. For the range-based protection, its computing overhead stays the same at different BER setups, but its benefit to accuracy increases with BER, because NLFs are usually placed after linear functions and also help to suppress the computing errors of linear functions especially when LB-ABFT fails to correct the computing errors at higher BER.

\begin{figure*}[!t]
\vspace{-0.5em}
\centering
\includegraphics[width=0.9\textwidth]{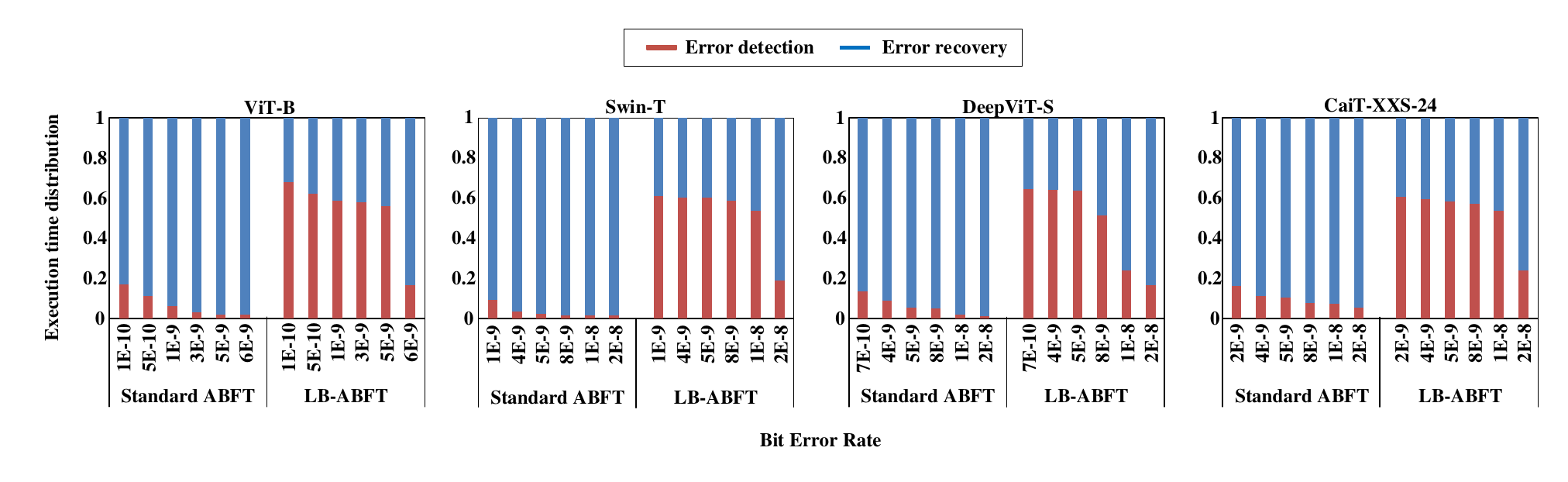}
\vspace{-1em}
\caption{The execution time distribution of standard ABFT and the proposed LB-ABFT under different bit error rate setups.}
\vspace{-0.5em}
\label{fig:check_recovery_time}
\end{figure*}

\begin{figure*}[!t]
\centering
\includegraphics[width=0.9\textwidth]{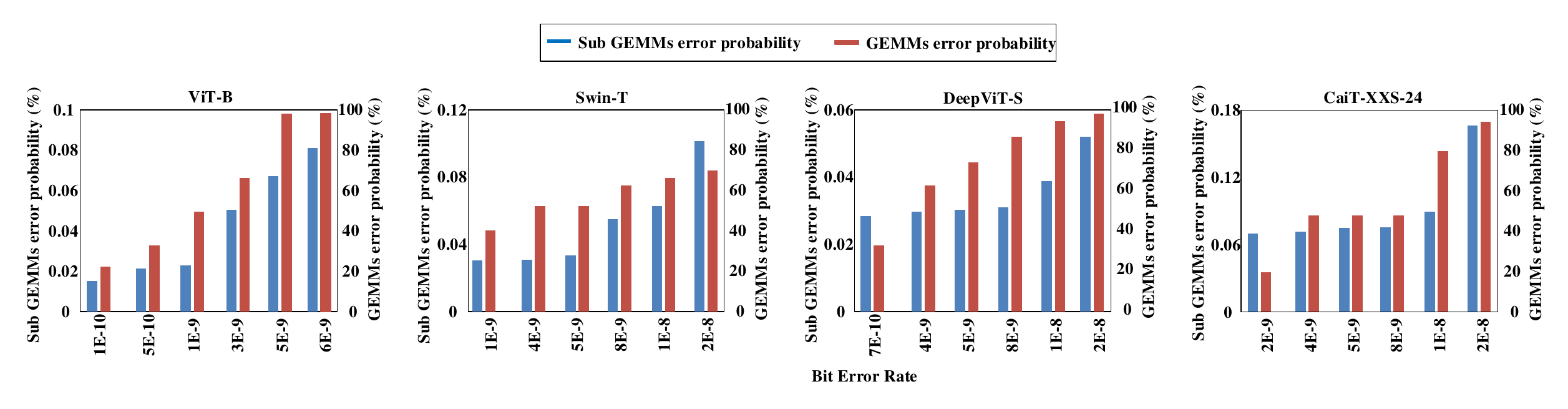}
\vspace{-1em}
\caption{The average error probability of sub GEMMs in LB-ABFT and the average error probability of GEMMs in standard ABFT, which directly reveals the invocation probability of error recovery procedures of LB-ABFT and standard ABFT under different BER setups.}
\vspace{-1em}
\label{fig:p_error_recovery}
\end{figure*}

To highlight the advantages of the proposed LB-ABFT, we compare it with several state-of-the-art fault-tolerant methods from the perspective of both model accuracy and computing overhead in Figure \ref{fig:pro_linear_acc} and Figure \ref{fig:pro_linear_flops}, respectively. \cite{R2F2021TVLSI} \cite{sabih2021fault} represent typical layer-wise TMR approaches. Despite the different application scenarios of ABFT, \cite{zhao2020ft} \cite{safarpour2021low, roffe2020evaluation, sharif2023efficient, li2022efficient, zamani2019greenmm, liu2018fault} essentially utilize standard ABFT approaches. They conduct ABFT with the granularity of GEMMs and can only recover from a single computing error. Particularly, we want to emphasize that standard ABFT is widely utilized and can still be considered to be one of the state-of-the-art algorithm-based fault tolerance approaches. \cite{kosaian2021arithmetic} is an intensity-guided ABFT and suppose it hides the error detection overhead within GPU idle time slot perfectly. It also recovers from a single computing error. According to the comparison, we notice that LB-ABFT achieves comparable accuracy with layer-wise TMR, but the computing overhead is much smaller. When compared with standard ABFT and intensity-guided ABFT, LB-ABFT requires much less multiplication and achieves higher model accuracy under all the different BER setups. The main reason can be attributed to two aspects. On the one hand, LB-ABFT divides larger GEMM into multiple sub blocks and spreads random bit-flip errors across the different sub blocks, which greatly improves the success rate of the LB-ABFT error recovery procedures. Thereby, the resulting model accuracy can be improved. On the other hand, some of the sub blocks are fault-free, and LB-ABFT only performs error recovery on faulty sub GEMMs instead of the whole GEMM. Hence, this approach reduces the total computing overhead. While intensity-guided ABFT shows less addition overhead than LB-ABFT because of the ideal error detection hiding mechanism, the optimization can be potentially orthogonal to this work.

In order to gain insight of the proposed LB-ABFT, we also present the execution time distribution of both LB-ABFT and standard ABFT deployed on GPUs in Figure \ref{fig:check_recovery_time}. It can be observed that the execution time distribution varies substantially. The error recovery procedure dominates the execution time for standard ABFT while the proportion of the error detection procedure is much higher in LB-ABFT. The main reason can be attributed to the fine granularity of error detection and error recovery in LB-ABFT. Basically, many sub GEMMs are actually fault-free and they will not invoke the time-consuming error recovery procedure. To further confirm the above analysis, we present the average error probability of sub GEMMs in LB-ABFT and the average error probability of GEMMs in standard ABFT in Figure \ref{fig:p_error_recovery}, which directly reveals the invocation probability of error recovery procedures of LB-ABFT and standard ABFT under different BER setups. As shown in Figure \ref{fig:p_error_recovery}, the error probability of sub GEMMs is very low in general and it is much lower than that of the standard ABFT without blocking. The higher error probability of GEMMs in standard ABFT will not be a problem at lower BER because the overall computing overhead is dominated by the error detection overhead, but it becomes expensive at higher BER when the error recovery procedure takes up the majority of the ABFT overhead as shown in Figure \ref{fig:check_recovery_time}.

\section{Conclusion}
In this paper, we perform comprehensive reliability analysis of ViTs at different granularities including models, layers, modules, and patches, and reveal the unique reliability features. Based on the analysis, we proposed LB-ABFT approach to adjust blocking size to fit the reliability requirements of different ViTs linear functions and BER setups, and applied a range-based approach to protect NLFs that are generally overlooked. According to our experiments, the proposed LB-ABFT achieves great advantages on both accuracy and computing overhead over the standard ABFT at various BER setups. The hybrid approach shows significant accuracy improvement and can almost fully recover the accuracy even when the accuracy drops by 50\%.

\bibliographystyle{ieeetr}
\bibliography{ref}

\vfill
\end{document}